\documentclass{aastex631}

\usepackage{hyperref}
\usepackage{mathrsfs}

\newcommand\bth{\frac{B_x}{|\vec{B}|}}
\newcommand\vo{\bar{v}_{{\rm out}}}
\newcommand\erf{{\rm{erf}}}
\newcommand\erfArg{\bth \frac{\vo}{\sqrt{\beta_i}}}
\newcommand\erfA{{\rm{erf}}\left( \bth \frac{\vo}{\sqrt{\beta_i}}\right)}
\newcommand\dhybridr{{\it dHybridR}}
\newcommand\bit{{\beta_{i, {\rm tot}}}}
\newcommand\bi{{\beta_{i}}}

\newcommand{\snowman}{\chi}

\usepackage{mathtools}
\usepackage{amsfonts}

\begin{document}

\title{Suppression of Collisionless Magnetic Reconnection in the High Ion $\beta$, Strong Guide Field Limit}

\author{Carlos A. Giai}
\affiliation{Institute for Astronomy, University of Hawai`i, Honolulu, HI, 96822, USA}
\affiliation{Now at Southwest Research Institute, San Antonio, TX 78238, USA}
\affiliation{Department of Physics and Astronomy, University of Texas at San Antonio, San Antonio, TX, 78238, USA}

\author[0000-0002-2160-7288]{Colby C. Haggerty}
\affiliation{Institute for Astronomy, University of Hawai`i, Honolulu, HI 96822, USA}

\author[0000-0003-1861-4767]{Michael A. Shay}
\affiliation{Bartol Research Institute, Department of Physics \& Astronomy, University of Delaware, Newark, DE, 19716, USA }

\author[0000-0002-5938-1050]{Paul A Cassak}
\affiliation{Department of Physics and Astronomy and the Center for KINETIC Plasma Physics, West Virginia University, Morgantown, West Virginia, 26506, USA}

\begin{abstract}
In magnetic reconnection, the ion bulk outflow speed and ion heating have been shown to be set by the available reconnecting magnetic energy, i.e., the energy stored in the reconnecting magnetic field ($B_r$).
However, recent simulations, observations, and theoretical works have shown that the released magnetic energy is inhibited by upstream ion plasma beta $\beta_{i}$ -the relative ion thermal pressure normalized to magnetic pressure based on the reconnecting field- for antiparallel magnetic field configurations.
Using kinetic theory and hybrid particle-in-cell simulations, we investigate the effects of $\beta_{i}$ on guide field reconnection.
While previous works have suggested that guide field reconnection is uninfluenced by $\beta_{i}$, we demonstrate that the reconnection process is modified and the outflow is reduced for sufficiently large $\beta_{i} > B_g^2/(B_r^2 + B_g^2)$.
We develop a theoretical framework that shows that this reduction is consistent with an enhanced exhaust pressure gradient, which reduces the outflow speed as  $v_0 \propto 1/\sqrt{\beta_{i}}$.
These results apply to systems in which guide field reconnection is embedded in hot plasmas, such as reconnection at the boundary of eddies in fully developed turbulence like the solar wind or the magnetosheath as well as downstream of shocks such as the heliosheath or the mergers of galaxy clusters.
\end{abstract}

\section{Introduction} \label{sec:intro}
Magnetic reconnection is a fundamental plasma process that rapidly converts energy stored in magnetic fields into kinetic energy in the form of heated and accelerated plasma. 
Reconnection is believed to be a potentially important phenomena in disparate plasma environments, from laboratory, to heliospheric to astrophysical systems; from solar/stellar flares, to substorm onset at the magnetotail, to impulsive high energy emission from astrophysical objects, magnetic reconnection is frequently invoked to explain the rapid generation of energetic ions and electrons \citep{dungey61,parker63,angelopoulos+08,zweibel+09,yamada+97}.
The theory of how reconnection heats and accelerates these particles in environments with infrequent collisions was the subject of numerous studies (e.g., \cite{hoshino+01,Drake+06,le+12,dahlin+14,yoo+14,phan+13,shay+14,haggerty+15,rowan+17}).
Recent works have connected ion energization in collisionless reconnection to a Fermi-like process in which the reconnection outflow jets, accelerated by the contracting magnetic field lines accelerates ions, producing counter streaming populations and increased heating \citep{Fermi49,Drake+06,drake+09a,Haggerty+16,haggerty+17,zhang+21}.
This energization depends on the reconnection outflow speed, which emphasizes the importance of understanding the asymptotic outflow value.

The jets produced by reconnection are a hallmark feature and are consistently identified in simulations, observations and in laboratories \citep{paschmann+79,sonnerup+81,sato+79}.
Historically, the outflow speed has been predicted to reach the Alfv\'en speed based on the sheared\footnote{In this work we will refer to this as the reconnecting field, i.e., the component of the magnetic field that changes direction. This is in contrast to the guide field component which remains constant across the layer. We refer to configuration where there is no guide field component as ``anti-parallel'' reconnection, which is equivalent to having a shear angle of $180^{\circ}$.} component of the magnetic field\citep{parker63} ($B/\sqrt{4\pi\rho}$, where $B$ is the magnetic field and $\rho$ is the mass density based on the inflowing plasma).
However, recent works have shown that in collisionless, anti-parallel reconnection, the outflow speed is consistently less than the standard prediction of the inflowing Alfv\'en speed; this suppression of the outflow speed and its reconnection rate has been connected to the ion heating in the exhaust \citep{liu+12,haggerty+18,li+21a}.

This prior research on ion thermal impacts on reconnection primarily focused on anti-parallel reconnection, which demonstrated a reduction in the outflow jets for low ion beta\footnote{$\bi = 2n_i T_i/(B_r^2/4\pi)$, where $n_i$ is the ion number density, $T_i$ is the ion temperature or second moment of the ion distribution function in units of energy, and $B_r$ is the reconnecting component of the magnetic field.}, $\bi$, reconnection, and showed a greater reduction for increasing values of $\bi$.
It was suggested that this effect and the associated reduction in outflow speed would not occur for reconnection with a sufficiently large guide field ($B_g > 0.4 B_r$ or a  shear angle between the magnetic field on across the current sheet less than $\sim 135^{\circ}$) \citep{haggerty+18}.
\cite{haggerty+18} connected the reduction of the reconnection outflow velocity to the increased exhaust temperature and \cite{li+21a} demonstrated that the reduction was directly associated with the exhaust pressure gradient that opposed the outflow direction.
Both of these works predicted that in the anti-parallel limit the outflow velocity should be less than the Alfv\'en speed even in the low $\bi$ limit ($v_{out}\sim 0.77v_A$ and $0.43 v_A$ in \cite{haggerty+18} and \cite{li+21a} respectively).
\cite{haggerty+18} demonstrated that when a guide field was present, the outflow speed reached the upstream Alfv\'en speed based on the reconnecting field.
However, this prior work focused mainly on simulations with relatively small ion beta values when including the guide field, with a maximum value of $\bit < 0.5$ (where $\bit = 8 \pi n_i T_i/(B_r^2 + B_g^2)$, which we refer to as the ``total'' ion beta throughout this work).
In this current work we show that while this trend holds true for lower $\bi$ guide field simulations, it breaks for sufficiently high values of $\bi$.
Thus the large $B_g$, large $\bit$ region of reconnection parameter space has yet to be accurate described.
These configurations are likely relevant to a number of different hot systems including the outer heliosphere or in the intracluster medium (ICM) of galaxy clusters, systems in which turbulence and correspondingly guide field reconnection is expected to occur. 
Recognizing this, our study examines high ion $\bi$, guide field reconnection, aiming to provide insights into a broader range of astrophysical phenomena.

To explore this regime, we present results from hybrid particle-in-cell simulations of guide field reconnection with increasing total ion beta. We find that for sufficiently large values of $\bit$, reconnection is modified and the outflow speed is reduced.
The results in this manuscript are organized in the following way:
Sec.~\ref{sec:sims} discusses the hybrid code, the simulation set up and the survey of simulations performed, Sec.~\ref{sec:results} presents the results from the simulations and highlights particular simulations to identify some of the key physics, and Sec.~\ref{sec:theory}, supplemented by the Appendix, develops a theoretical prediction for the effects of increased ion $\bi$ on guide field reconnection which is shown to match well with the simulation results.
Finally, in Sec.~\ref{sec:fin}, we discuss implications for these results and conclusions. 

\section{Hybrid PIC simulations}\label{sec:sims}

To study the effects of increasing inflowing ion temperature on guide field reconnection,
we perform a survey of simulations using \dhybridr, a kinetic hybrid particle-in-cell (PIC) code \citep{gargate+07,haggerty+19a}. \dhybridr{} treats ions as macro particles that follow trajectories in phase space defined by the Lorentz force law. Electrons are treated as a massless charge neutralizing fluid who’s dynamics are taken to satisfy Ohm's law,
\begin{equation}
    \vec{E} = -\frac{\vec{u}_i}{c}\times \vec{B} + \frac{\vec{J}}{enc}\times \vec{B} - \frac{1}{en}\vec{\nabla}P_e,
\end{equation}
where $\vec{E}$ is the electric field, $\vec{u}_i$ is the bulk ion velocity, $c$ is the speed of light, $\vec{B}$ is the magnetic field, $\vec{J}$ is the current density, $e$ is the ion charge, $n = n_i = n_e$ is the number density of ions taken to be equal to the electron number density $n_e$, and $\vec{\nabla}P_e$ is the gradient of the electron pressure. 
The current density is set by the spatial variation of the magnetic field using Ampere's law neglecting the displacement current.
The electron pressure is determined using a polytropic equation of state, $P_e \propto n^\gamma$, where $\gamma = 5/3$.
Note that this version of Ohm's law omits the electron inertia term.
While \dhybridr{} retains the relativistic dynamics of the ions \citep{haggerty+20,caprioli+20}, the speed of light is set sufficiently high in these simulations so that relativistic effects can be neglected.

The simulations are 2.5D (2D in real  and 3D in velocity space) on rectangular domains in the $x-y$ plane.
The magnetic field $B_{0}$ is the asymptotic reconnecting magnetic field, number density $n_{0}$ is normalized to an arbitrary characteristic value, and all other simulation values are normalized to them through the following plasma parameters: lengths  are normalized to the ion inertial length  $ d_{i}=c/\omega_{pi}$ where $\omega_{pi}$ is the ion plasma frequency $\sqrt{\frac{4 \pi n_{0} e^2}{m_i}}$. 
Time is normalized to the ion cyclotron time  $ \Omega^{-1}_{ci0} = m_{i}c/ e B_{0}$. Speeds are normalized to the Alfv\'en speed $ {V_{A}} = d_i\Omega_{ci} = \sqrt{B_{0}^{2} / 4 \pi m_{i} n_{0}}$.
Electric fields are normalized to $E_{0}=V_{A} B_{0}/c$ and temperatures to $T_{0}=m_{i}V^{2}_{A}$.

The simulation box has doubly periodic boundaries, and the coordinates $\hat{x}$ and $\hat{y}$ correspond to the outflow and inflow, respectively.
The simulations are $L_x \times L_y = 400 d_i\times 200 d_i$ with 2 grid points per $d_i$, a time step of $\Delta t = 0.01 \Omega^{-1}_{ci0}$, and a reduced speed of light of $c = 20 V_A$.
The simulations are initialized with two force-free current sheets with magnetic profile 
$\textbf{B} = B_{r}(\tanh((y-0.25L_{y})/\lambda) - \tanh((y-0.75L_{y})/\lambda)) \hat{\bf x}
+ (B_g + B_{\rm ff}(y)) \hat{\bf z}$
where, $B_{r}$ is the strength of the reconnecting magnetic field (set to $1B_0$ for every simulation), $B_g$ is the strength of the out-of-plane (or guide) field, $B_{\rm ff}(y)$ varies to maintain pressure balance across the current sheet and is zero outside of the current sheet, and $\lambda$ is the half-thickness of the current sheet. 
The ions are initialized with 100 particles-per-cell (PPC), with some select simulations rerun with a significantly higher PPC of 10,000 for better counting statistics and for smoother higher order moments (denoted by the simulations with a $\dagger$ in Table~\ref{tab:sims}).
The ions are initialized as a Maxwell–Boltzmann with constant density and temperature across the current sheet as well as no initial bulk flow. The initial ion temperature along with the guide field is varied between simulations as described in Table~\ref{tab:sims}. The electron temperature is chosen to be 0.1 for all simulations.


\begin{table}
\centering
\begin{tabular}{|c|c|c|c|c|}
\hline
Sim Number & $T_i$ ($m_{i}V^2_{A0}$) & $B_g (B_0)$ & $\beta_i$ & $\beta_{i,{\rm tot}}$ \\
\hline
\hline1  & 1.00 & 0.00 & 2.00 & 2.00 \\ 
\hline
3  & 1.00 & 1.00 & 2.00 & 1.00 \\ 
\hline
6$^\dagger$  & 16.00 & 2.00 & 32.00 & 6.40 \\ 
\hline
7$^\dagger$  & 2.00 & 0.00 & 4.00 & 4.00 \\ 
\hline
8$^\dagger$  & 2.00 & 1.00 & 4.00 & 2.00 \\ 
\hline
9  & 4.00 & 0.00 & 8.00 & 8.00 \\ 
\hline
10$^\dagger$  & 4.00 & 1.00 & 8.00 & 4.00 \\ 
\hline
11$^\dagger$  & 6.00 & 0.00 & 12.00 & 12.00 \\ 
\hline
12$^\dagger$  & 6.00 & 1.00 & 12.00 & 6.00 \\ 
\hline
13  & 0.10 & 0.00 & 0.20 & 0.20 \\ 
\hline
14  & 0.25 & 0.00 & 0.50 & 0.50 \\ 
\hline
15  & 0.50 & 0.00 & 1.00 & 1.00 \\ 
\hline
16$^\dagger$  & 2.50 & 2.00 & 5.00 & 1.00 \\ 
\hline
17$^\dagger$  & 2.00 & 2.00 & 4.00 & 0.80 \\ 
\hline
19  & 1.00 & 2.00 & 2.00 & 0.40 \\ 
\hline
20  & 0.30 & 1.00 & 0.60 & 0.30 \\ 
\hline
21$^\dagger$  & 6.25 & 2.00 & 12.50 & 2.50 \\ 
\hline
22  & 0.62 & 2.00 & 1.24 & 0.25 \\ 
\hline
\end{tabular}
\caption{Table of simulation parameters showcasing a range of guide field strengths and ion temperatures. All simulations are conducted in a simulation box with dimensions $L_x \times L_y = 400d_i \times 200d_i$. The grid resolution is set at 2 grid points per $d_i$ with a time step of $\Delta t = 0.01\Omega^{-1}_{ci0}$ and a reduced speed of light $c = 20v_A$. The PPC count was 100 for standard simulations, while those marked with a dagger $^\dagger$ were additionally run with a PPC of 10,000 to reduce counting statistics noise. The simulations consistently used a reconnecting magnetic field strength of $1B_0$, with the out-of-plane (guide) field strength, $B_g$, and ion temperature, $T_i$, varying across different runs.}
\label{tab:sims}
\end{table}

\section{Results}\label{sec:results}
\subsection{Overview}
\begin{figure}[ht]
    \centering
    \includegraphics[width=.99\columnwidth]{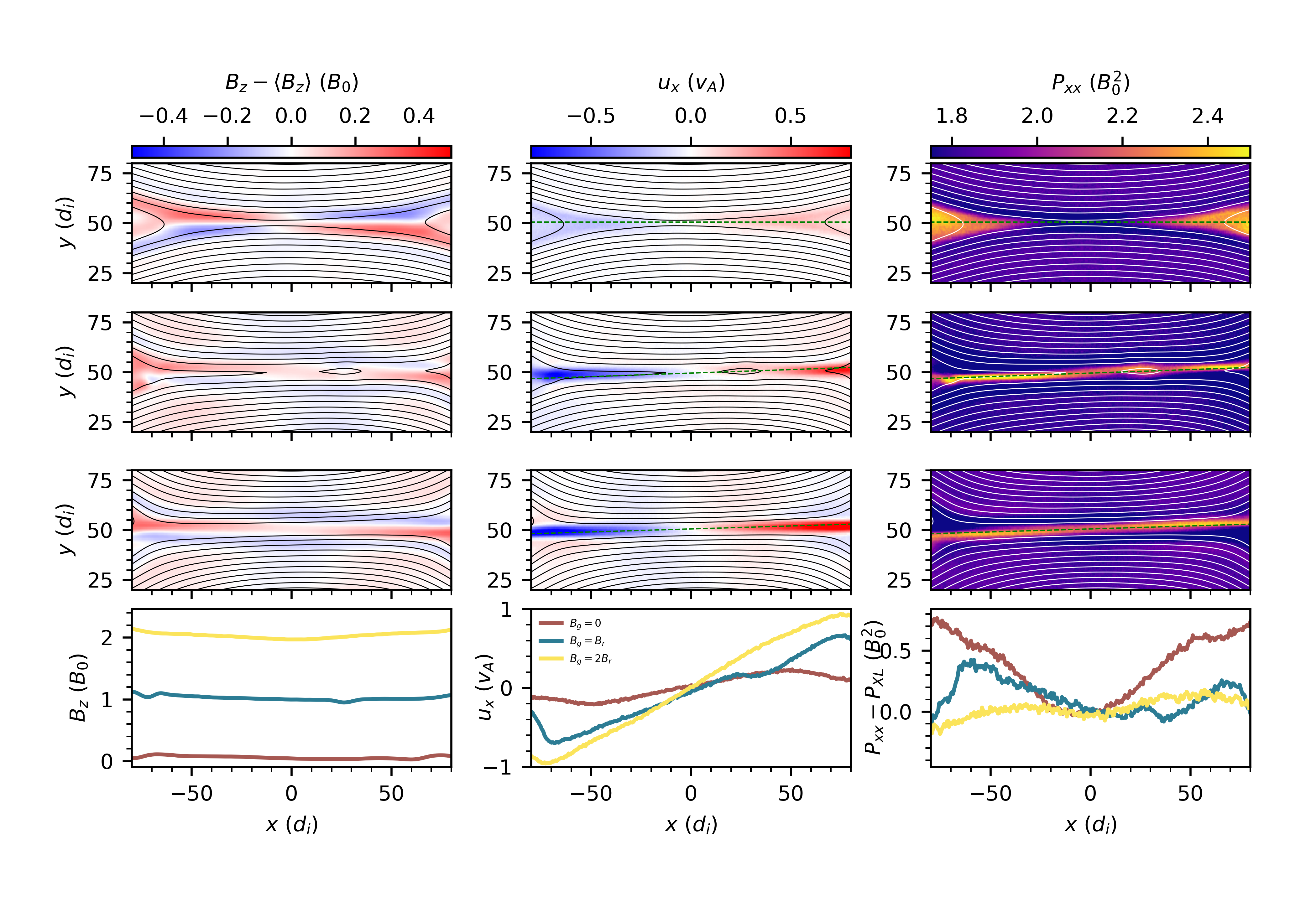}
    \caption{Comparison of $\bi = 4$ simulations with increasing guide field strength. The first three rows correspond to different simulations: anti-parallel reconnection (top row, run $7^\dagger$), moderate guide field (middle row, $8^\dagger$), and strong guide field (bottom row, $17^\dagger$) displaying a subset of each simulation. From left to right the columns show the deviation from the mean guide field strength $B_z - \left< B_z \right>$, outflow speed $u_x$, and the $P_{ixx}$ component of the pressure. The fourth row shows a 1D cut along the exhaust (mostly along x, but slightly tilted to align with the outflow jet and pressure enhancement) for each of the simulations ($B_g = 0, 1, 2 B_r$ for brown, blue and yellow lines, respectively). The cuts show that the reduction in outflow speed and a pressure gradient between the x-line and the exhaust are anti-correlated with increasing guide field.}
    \label{fig:overview}
\end{figure}

The results of the simulation survey reproduce many of the trends found in anti-parallel reconnection \citep{drake+09a,haggerty+18,li+21a}; for relatively low $\bi$, the reconnection outflow speed is decreased in anti-parallel systems, yet reaches approximately the Alfv\'en speed based on the reconnecting field for the strong guide field case ($B_g \gtrsim B_r$).
This trend is demonstrated in Fig.~\ref{fig:overview}, which compares three different simulations across the first, second and third rows for $B_g = 0,1,2B_r$, respectively, with all simulations initialized with $\bi = 4$.
The first column shows the deviation from the mean of the out-of-plane magnetic field strength, the second column shows the ion bulk velocity in the $x$-direction (showing the reconnecting outflow jets) and the last column shows the $xx$ component of the ion pressure tensor $P_{xx}$.
The 2D bulk velocity plots show that the outflow velocity is sub-Alfv\'enic for the anti-parallel simulation, but increases and saturates near the Alfv\'en speed for the stronger guide field simulations.
The final column shows that the $xx$ component of the ion pressure in the exhaust is anti-correlated with the outflow speed, i.e., the change in the ion pressure between the exhaust and x-line is much larger in the anti-parallel simulation compared to the guide field simulation.
This point is further quantified in the bottom row which shows 1D cuts through the exhaust\footnote{Note that the cut is not exactly horizontal for the guide field simulations, while it is mostly along $x$ is has been slightly tilted in $y$ to align with the direction of the outflow jet and the pressure enhancement. From \ref{fig:overview}, it is clear that the $xx$ component of the pressure is tilted similar to the outflow} for each of the simulation (color coded with brown, blue and yellow to $B_g = 0,1,\ 2B_r$ respectively).
The bottom middle panel shows that as the guide field increases, the outflow speed in the exhaust increases and asymptotes at the Alfv\'en speed for the $B_g = 2B_r$ simulation.
Similarly, the bottom right panel shows that this is anti-correlated with the change in $P_{ixx}$ between the exhaust and the x-line.

These results are consistent with the theory put forward by \cite{li+21a}, which linked the reduction in outflow speed to the back pressure gradient created by the ions heated by reconnection.
This demonstrates that the hybrid simulations used in this work reproduce the results identified in fully kinetic PIC simulations as well as observations \citep{haggerty+18,li+21a};
this reaffirms that ion scale physics is controlling this process and the validity of the hybrid method for addressing this problem.

\subsection{Increasing Total $\beta$}
\begin{figure}[ht]
    \centering
    \includegraphics[width=0.8\columnwidth]{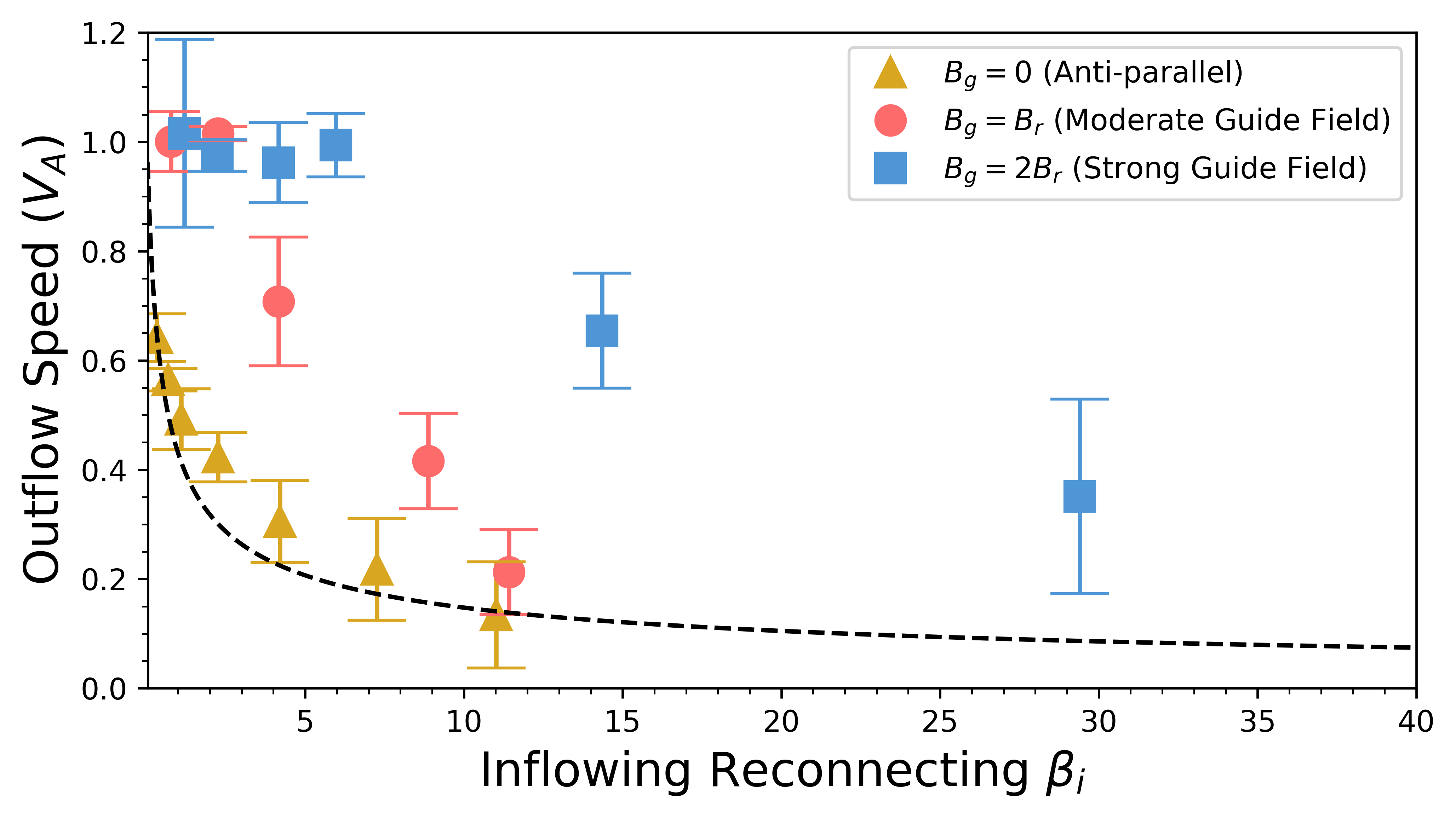}
    \caption{Average outflow speeds from reconnection simulations plotted against the reconnecting beta $\bi$ for various initial guide field values ($B_{g} = 0B_r$, gold triangle, $B_{g} = 1B_r$, pink circle, and $B_{g} = 2B_r$, blue square). The outflow speed is reduced for sufficiently high inflowing ion temperatures, even in the limit of a strong guide field. The black dashed line represents the prediction for anti-parallel reconnection as proposed by \cite{haggerty+18}, serving as a benchmark for comparison.}
    \label{fig:reconn_beta_outflow}
\end{figure}

While the low $\bi$ simulations show results consistent with previous works, the observed behavior changes for larger values of $\bi$ in the guide field simulations. We find that as $\bi$ increases, even in the guide field simulations, the outflow speed is reduced. This is displayed in Fig.~\ref{fig:reconn_beta_outflow}, which shows a scatter plot of average ion outflow vs inflowing reconnecting beta for all of the simulations.
The averages are determined by taking a cut along $y$ through one of the four exhausts roughly $70 d_i$ downstream of the x-line and the upstream and downstream averaging regions are determined by eye.
The points are color coded with the strength of the initial guide field (gold, pink and blue for $B_g = 0,\,1,\,2 B_r$, respectively).
The anti-parallel simulations show the same behavior as previous works (with the black dashed line the empirical prediction from \cite{haggerty+18}); as $\bi$ increases, the outflow speed decreases. Similarly, for the guide field simulations, we find that as $\bi$ increases, eventually the outflow speed drops in a manner comparable to the anti-parallel simulations.
From this plot, there is an apparent $\bi$ threshold, below which reconnection is unaffected by thermal variation (as demonstrated by all the pink and blue points with an outflow speed $\sim V_A$).
Furthermore, this threshold has a dependence on the guide field strength, as the cut-off corresponds to larger $\bi$ values for stronger guide fields.

\begin{figure}[ht]
    \centering
    \includegraphics[width=0.8\columnwidth]{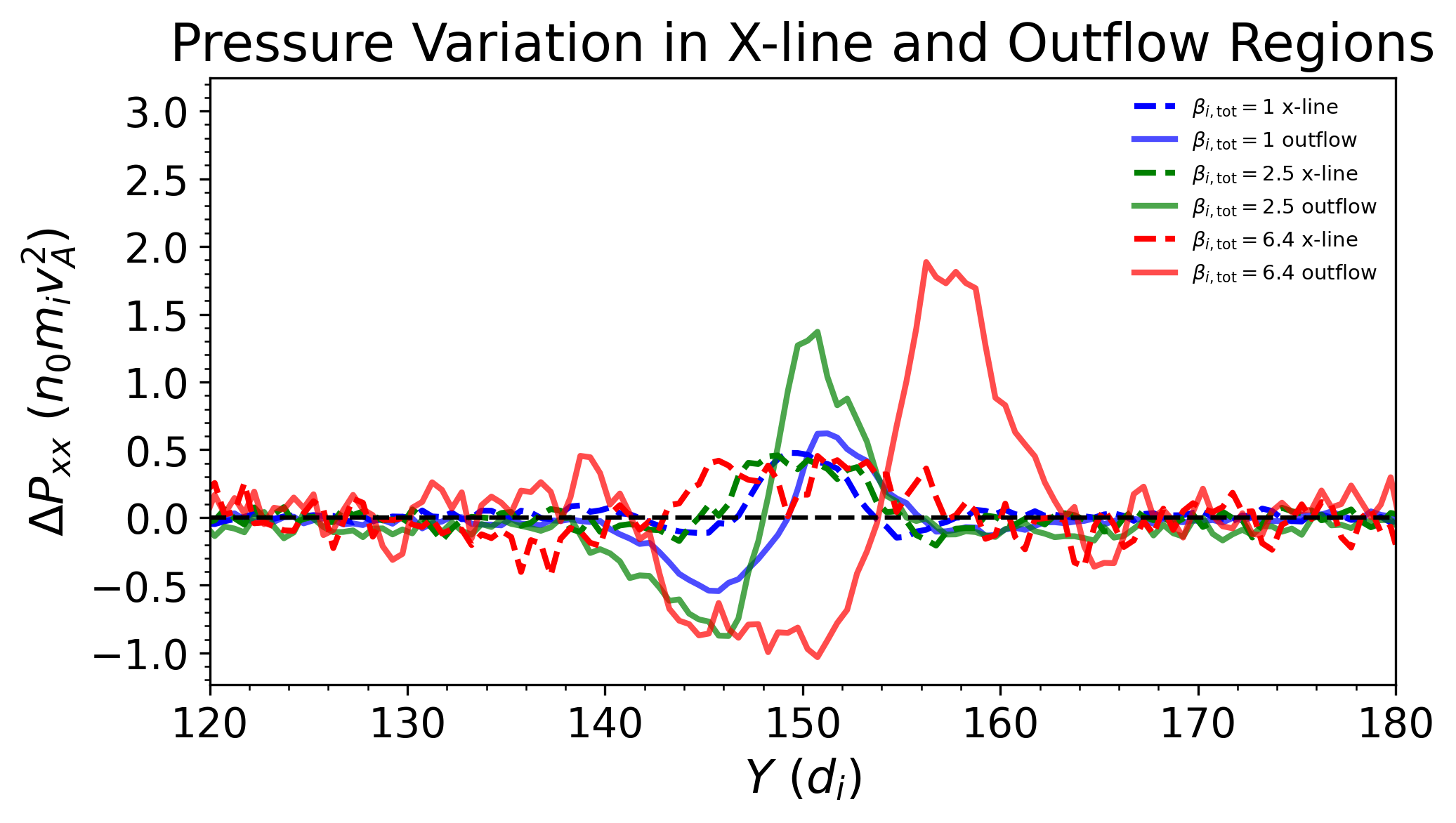}
    \caption{Pressure profiles from simulations $16^\dagger$, $21^\dagger$, and $6^\dagger$, each represented by a unique color: blue for $\bit = 1$, green for $\bit = 2.5$, and red for $\bit = 6.4$. All simulations share a common guide field of $B_{g} = 2B_{r}$. The solid lines denote the cuts along the exhaust, while the dashed lines signify the x-line cuts. A key observation is the amplifying disparity between the peak values in these regions with increasing total initial beta $\bit$. The exhaust pressure increasing along with the upstream temperature, while the pressure at the x-line remains roughly constant, corresponding to a pressure gradient opposing the outflow.}
    \label{fig:deltaP}
\end{figure}

Ultimately we find that the physics of the outflow reduction in anti-parallel reconnection is analagous to the guide field case as well; namely that for large $\bi$, a large ion exhaust pressure gradient develops and reduces the outflow speed \cite{li+21a}.
This is shown in Fig.~\ref{fig:deltaP}, which shows cuts along $y$ of the $xx$ component of the ion pressure tensor, through both the x-line (dashed line) and outflow region (solid line) for three strong guide field simulations ($B_g = 2B_r$) with increasing inflowing temperature ($T_i = 2.5, 6.25,\ 16.0 m_iV_A^2$ for the blue, green and red lines, respectively).
For the low $\bit$ simulation (blue), the pressure at the x-line and the exhaust remains small and roughly constant, meaning that there is no pressure gradient to reduce the outflow speed.
As $\bit$ increases (green and red), the peak values in the $P_{ixx}$ profile become more pronounced, while the pressure at the x-line remains roughly unchanged;
this indicates a substantial increase in the pressure gradient between the exhaust and the x-line, which is consistent with the reduced outflow speed identified in these simulations. 
This shows that increasing the inflow temperature increases the change in pressure in the exhaust, which enhances the thermal effects on the reconnection process, even in the presence of a guide field.

It is noteworthy that the result identified in the simulation (i.e., that $\Delta T_i \sim m_i v_{\rm out}$) differs from the typical prediction for ion heating which has a quadratic dependence with the outflow speed, $\Delta T_i \sim m_i v_{\rm out}^2$ \citep{Drake+06,phan+14,shay+14}.
This deviation in behavior is because not all of the upstream ions reach the exhaust because of the larger thermal speed for ions flowing away from the reconnection region. The resulting exhaust distribution is then made up of hotter segments of inflowing ions and form a non-Maxwellian distribution \citep{drake+14,li+21a}; this is discussed further in Sec.~\ref{sec:theory}.

Finally, from the simulations we find that the reduction in outflow is well correlated with {\it total} ion $\bit$, opposed to the reconnecting $\bi$, where $\bit$ is based on the magnetic pressure of both the reconnecting and guide field.
This is shown in Fig.~\ref{fig:total_beta_outflow} which is nearly the same plot as Fig.~\ref{fig:reconn_beta_outflow}, except that the x-axis is in terms of $\bit$.
Plotted in this way, the variation due to guide field strength vanishes, and the pink and blue points follow roughly the same relation as the gold points.
Furthermore, we find that the threshold between Alfv\'enic and sub-Alfv\'enic behaviors lies at a value of approximately $\bit = 1$.
For $\bit < 1$ (the pink shaded region), the outflow speeds converge on the Alfv\'en speed, for each guide field strength.

It is notable that we identified a sharp transition around $\bit \sim 1$, as it is the boundary for other processes in reconnection. In guide field reconnection, the physics of the ion diffusion region can be split between whistler-mediated ($\bit > 1$) and kinetic Alfv\'en wave (KAW)-mediated ($\bit < 1$)\citep{rogers+01}.
The reconnection is characterized in the two limits by the self-generated out-of-plane magnetic fields with the whistler physics by a symmetric quadrupolar $B_z$ perturbation \citep{tenbarge+14,sharma+19} and the KAW physics having more complex $B_z$ structure\citep{rogers+03}.
The idea that ions are demagnetizing above this threshold is consistent with these results as the transition from KAW to whistler-like physics occurs due to the changing response of the ions.

\section{Theory}\label{sec:theory}
\begin{figure}[ht]
    \centering
    \includegraphics[width=0.8\columnwidth]{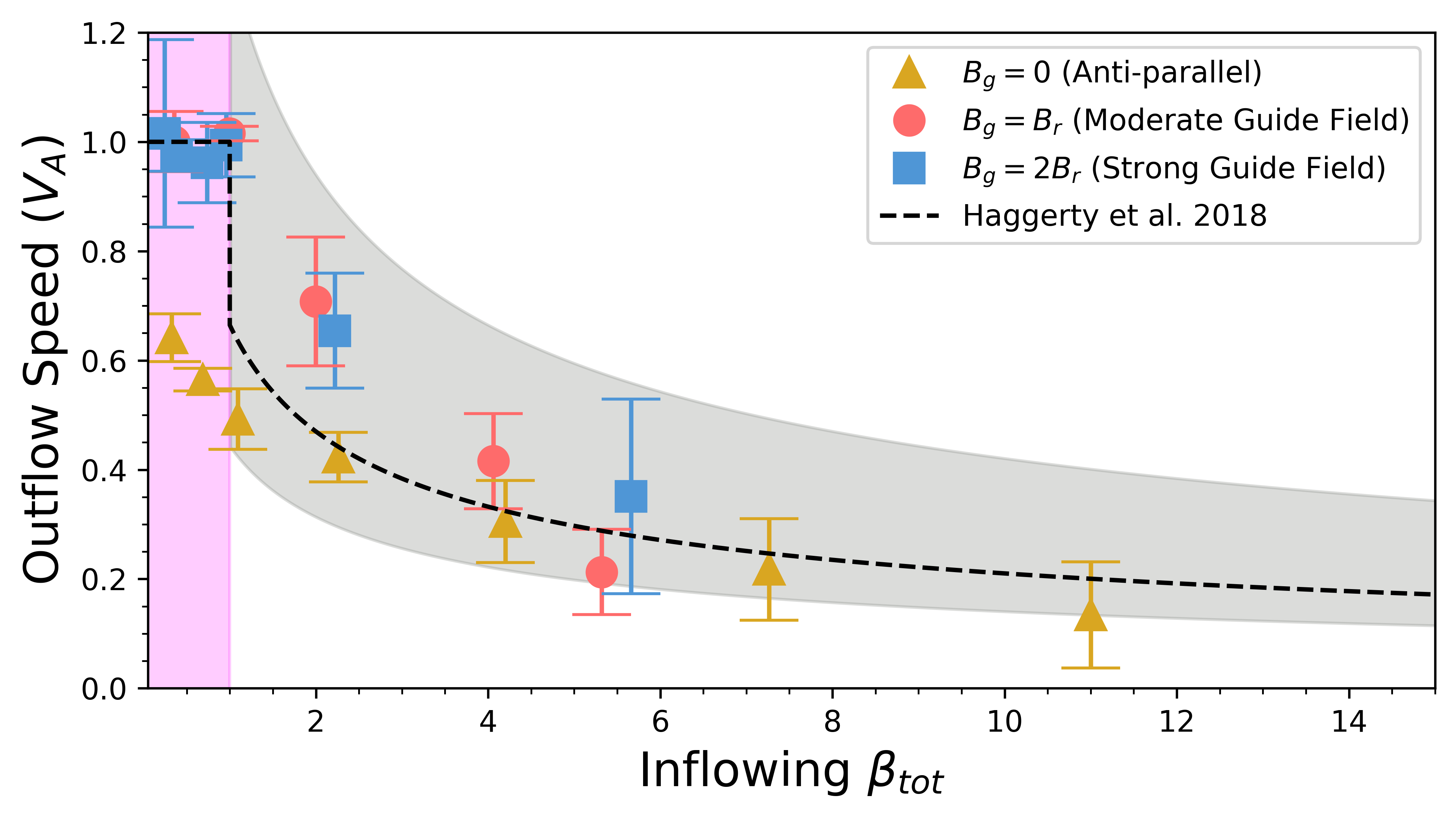}
    \caption{Simulation survey of the outflow dependence on total beta $\bit$. The vertical axis represents outflow speed, while the horizontal axis shows the total inflowing beta. Different shapes correspond to varying initial guide field values. The gray area illustrates the range of predictions from this work depending on the anisotropy of the ion pressure in the exhaust. The lower boundary is corresponds to all of the exhaust heating occurring in the $xx$ of the ion temperature tensor, while the upper bound is if the exhaust heating is isotropic; the black dashed line corresponds to the ion pressure anisotropy inferred from simulations.
}
    \label{fig:total_beta_outflow}
\end{figure}

To understand the behavior observed in the simulations we need a theory to include the presence of a guide field.
Fig.~\ref{fig:reconn_beta_outflow} demonstrates that for sufficiently large $\bi$, even guide field simulations have a reduced outflow.
This behavior is understood by considering the ion heating mechanism discussed in \cite{drake+09b}; for sufficiently large reconnecting $\bi$ values ($\bi \gtrapprox 0.2$), heating in the strong guide field limit is attributed to the counter streaming beams of ions inflowing parallel to magnetic field lines.
This heating is estimated by boosting into the reference frame moving with the outflowing plasma and then determining the temperature increment for the counter-streaming populations,
\begin{equation}
    \Delta T_{i\parallel} = m_iv_{\rm out}^2\frac{B_{x}^2}{|B|^2},\ \Delta T_{i\perp} = 0,
    \label{eq:jims_heating}
\end{equation}
where $\parallel$ and $\perp$ are relative to the local magnetic field direction and $|\vec{B}|^2 = B_g^2 + B_r^2$.
For guide field reconnection, this increased heating corresponds to the $T_{izz}$ term of the temperature tensor near the center of the exhaust.
The outflow is reduced by the pressure gradient along the outflow direction, and so only the $P_{ixx} = n_iT_{ixx}$ term should contribute.
This implies that the outflow speed should not be reduced in the strong guide field limit, consistent with the guide field simulations in the shaded violet region in Fig.~\ref{fig:total_beta_outflow}.
However, from Fig.~\ref{fig:total_beta_outflow}, this behavior does not hold for $\bit \gtrapprox 1$.

For the $\bit \gtrapprox 1$ regime, the decreased outflow velocity is consistent with an increase in the $xx$ component of the pressure tensor (as shown in bottom right panel of Fig.~\ref{fig:overview}, and Fig.~\ref{fig:deltaP}).
This implies that Eq.~\ref{eq:jims_heating} is incorrect for $\bit > 1$.
However, Eq.~\ref{eq:jims_heating} can be updated for this limit by including two additional considerations; the ion exhaust temperature increases with the inflowing ion temperature and the ions in the exhaust demagnetize and scatter as they pass through the center of the current sheet.

The demagnetization occurs around $\bit \gtrapprox 1$ because the ion gyroradius becomes approximately larger than the current layer ($r_g \gtrapprox c/\omega_{pi}$) \citep{shay+99}. This is discussed and demonstrated in Appendix~\ref{sec:apdx:CS}.
This scattering allows for ion heating along the outflow direction, rather than only the parallel direction as is typically predicted for the strong guide field case.
While one would expect that this scattering would simply cause the ion exhaust temperature to become isotropic, in the simulations we find a slight preferential heating in the $xx$ direction.

The increased ion exhaust heating due to larger values of $\bit$ occurs because of the larger fraction of ions that are able to ``out run'' the exhaust.
For larger temperature, ions with a large field-aligned velocity directed away from the exhaust move away from the reconnection region faster than the inflowing field lines can bring these ion into the exhaust.
Because of this, the ion exhaust is made up of two populations that have been accelerated by the reconnection process (see Fig.~3d in \cite{li+21a}), and these two effects contribute to increasing the ion exhaust pressure.

The theory of \cite{li+21a} can be updated to include a guide field and we show that in the guide field limit the reduction in the outflow speed depends on the total upstream ion plasma beta $\bit \equiv 8\pi n_i T_i/(B_r^2 + B_g^2)$.
We start with the force balance equation in the exhaust determined at the point between the x-line and the outer edge of the ion diffusion region (see \cite{li+21a}, Eq.~6 and Appendix~C for details, noting their different coordinate system convention).
\begin{equation}
\frac{n_2m_i v^2_{\rm out}}{2\Delta x} + \frac{B^2_y}{8\pi \Delta x} + \frac{\Delta P_{ixx}}{\Delta x} = \frac{B_y}{8\pi}\frac{\epsilon B_{xm}}{\Delta y},
\label{eq:PresBal_NoNorm}
\end{equation}
where $B_{xm}$ is measured at the inflow edge of the ion diffusion region, $\Delta x$ and $\Delta y$ are the width and height of the diffusion region respectively, $\epsilon$ is the firehose parameter, $n_2$ is the density in the exhaust, and $\Delta P_{ixx}$ is the difference in the $xx$ component of the pressure gradient between the x-line and exhaust.
This equation can be simplified by normalizing by $m_i n_0 V_A^2$ and using $B_y/B_{xm} \simeq \Delta y/\Delta x$ we find
\begin{equation}
    \frac{1}{2}\frac{n_2}{n_0} \bar{v}_{{\rm out}}^2 + \Delta \bar{P}_{ixx} + \frac{\bar{B}^2_{x}}{2}\left[ \left(\frac{\Delta y}{\Delta x} \right)^2 - \epsilon \right] \approx 0,
    \label{eq:PresBal}
\end{equation}
where $\bar{v}_{{\rm out}} = v_{\rm out}/V_A$, $\bar{B}_x = B_{xm}/B_{x0}$ and $\Delta \bar{P}_{ixx} = (P_{ixx} - P_{i,{\rm x-line}})/(B_{x0}^2/4\pi)$ and $P_{i,{\rm x-line}}$ is the $xx$ component of the ion pressure tensor at the x-line.
This equation is not modified by including a guide field, as the out-of-plane magnetic field does not contribute to the magnetic tension force.
This equation can be used to solve for the normalized outflow if the pressure can be described as a function of the outflow velocity.
Such an equation can be determined in the strong guide field limit by considering two cases.

In the low $\bit < 1$ regime, the heating due to reconnection is predicted to be anisotropic and as ions are expected to be preferentially accelerated parallel to field lines. This implies that $\Delta \bar{P}_{ixx} \sim 0$, as is demonstrated by the blue lines in Fig.~\ref{fig:deltaP}.
Additionally in the guide field limit, the density at both the x-line and the exhaust are comparable and so $n_2 \approx n_0$, motivated by the higher beta limit and verified in the simulations. 
Taking the diffusion region aspect ratio to be small ($\Delta y/\Delta x \ll 1$, $\bar{B}_x\rightarrow 1$ and $\epsilon \rightarrow 1$) we find
\begin{equation}
    \bar{v}_{\rm out} \approx 1 \Rightarrow v_0 \approx V_A,
\end{equation}
consistent with what is shown for $\bit < 1$ in Fig.~\ref{fig:total_beta_outflow}.

However, for a sufficiently large temperature, the ions demagnetize as they cross the current sheet, causing an increase in the $xx$ component of the pressure.
This transition occurs when the gyroradius becomes comparable to the current sheet width of roughly $\sim d_i$, or $r_g/d_i \gtrsim 1$ or equivalently $\sqrt{\bit} \gtrsim 1$.
In this higher beta limit, the exhaust pressure is set by two competing effects; first the ion heating will be enhanced by the increased upstream ion temperature as discussed in both \cite{drake+14} and \cite{li+21a}, and the heating is reduced by the guide field (as discussed in \cite{drake+09b}).
We combine these two effects to calculate the normalized exhaust pressure as 
\begin{equation}
\Delta \bar{P}_{ixx} =  \snowman  \left[ \left( \vo \bth \right )^2 + \left[ \left( \vo \bth \right )^2 + \frac{\bi}{2}\right ] \erfA
+ \vo \bth \sqrt{\frac{\bi}{\pi}}\exp{\left({-\erfArg}\right)} \right], \label{eq:Pnorm}
\end{equation}
where $\erf$ is the error function defined as $\erf(x) = \frac{2}{\sqrt{\pi}} \int_{0}^{x} e^{-x'^2} \, dx'$ and $\snowman$ is a dimensionless scattering parameter that is between 0 and 1. 
The $\snowman$ parameter is introduced because it is uncertain how much energy scattering transfers into the $T_{ixx}$ component.
As discussed above, if no scattering occurs $\snowman = 0$ and we recover the Alfv\'en speed for $v_{\rm out}$.
If all of the energy is channeled into the $xx$ component, then $\snowman = 1$, and if the exhaust temperature is isotropic then $\snowman = 1/3$ because the $xx$ component of the temperature only gets $1/3$ of the total energy that went into heating.
From the hybrid simulations we find the $xx$ component to be roughly a factor of 2 larger than the $yy$ components (with the $zz$ component remaining small), which implies that $\snowman \approx 2/3$.
In Eq.~\ref{eq:Pnorm}, the ratio of the reconnection field to the magnitude of the total magnetic field appears in the equation because the ion parallel velocity that increases the exhaust temperature is reduced when changing into the outflow's reference frame \citep{drake+09a}.
Note that this equation converges to the results of \citep{li+21a} in the anti-parallel limit (their Eq.~10).
This equation can be used in Eq.~\ref{eq:PresBal} to develop an equation whose roots provide the prediction for the outflow speed.

In the large $\bi$ limit we can simplify our expression for the pressure by approximating the error function by $\frac{2}{\sqrt{\pi}}\erfArg$ and taking the exponential as 1.
Substituting this into Eq.~\ref{eq:Pnorm} and keeping only the highest order terms in $\bi$ we can derive a rather simple relationship for the exhaust pressure
\begin{equation}
    \Delta P_{ixx} \approx 2\snowman \bth \sqrt{\frac{\bi}{\pi}} m_i n_0 v_{\rm out} V_A = 2\snowman \sqrt{\frac{\bit}{\pi}} m_i n_0 v_{\rm out} V_A,\label{eq:Psimp}
\end{equation}
which is consistent to the prediction of \cite{drake+14} up to a constant factor of order unity.
The increased pressure in the exhaust opposes the magnetic tension force and reduces the outflow speed.
Substituting Eq.~\ref{eq:Psimp} into Eq.~\ref{eq:PresBal} we can derive a prediction for the outflow speed in the high $\bi$ limit
\begin{equation}
    \vo = \frac{\epsilon}{4\snowman }\sqrt{\frac{\pi}{\bi}}\bth = 
    \frac{\epsilon}{4\snowman }\sqrt{\frac{\pi}{\bit}} \approx
     \frac{0.664}{\sqrt{\bit}},\label{eq:simpPredict}
\end{equation}
where the last step is evaluated in the limit where $\snowman = 2/3$ and $\epsilon = 1$.
This prediction shows that the outflow velocity should decrease with the {\it total} ion beta, rather than reconnecting ion beta, a prediction that is consistent with what is found in the simulations.
This prediction is shown as the black dashed line in Fig.~\ref{fig:total_beta_outflow}, and is found to be in excellent agreement with the simulation data. 
A notable result of this prediction is that the outflow speed depends on $\bit$, rather than $\bi$, which is why the guide field simulations roughly follow the same curve, regardless of guide field strength.
It should be noted that the prediction in Eq.~\ref{eq:simpPredict} depends on the fraction of heating that is channeled into the $xx$ component, as parameterized by $\snowman$.
While we determine this value with the simulations, we can identify sensible upper and lower bounds for this parameter as shown in shaded gray region in Fig.~\ref{fig:total_beta_outflow}. 
The lower corresponds to a scattering parameter of $\snowman = 1$, which would mean that the temperature increase is strictly in the $xx$ direction and the upper bound corresponds to a scattering parameter of $\snowman = 1/3$ which would correspond to isotropic heating\footnote{Note we do not include the bound for the scattering parameter going to zero, as the equation is derived in the limit where the pressure term is large compared to the energy density in the outflow.}.
The strong agreement between theory and simulation in Fig.~\ref{fig:total_beta_outflow} and the increasing pressure demonstrated in Fig.~\ref{fig:deltaP} shows that the reduction in the outflow velocity is consistent with magnetic tension force being opposed by the back pressure gradient of the heated ions in the exhaust.
This further supports a growing consensus that thermal contributions to the reconnection process can be significant, and that the outflow speed is frequently sub-Alfv\'enic, especially when the reconnecting plasma is hot.

\section{Discussion and Conclusion}\label{sec:fin}
Our findings show that the outflow speed in guide field reconnection undergoes a reduction akin to that observed in antiparallel reconnection configurations when subjected to sufficiently high ion temperatures. This highlights the role of upstream plasma beta in modulating outflow speeds during magnetic reconnection, encompassing both antiparallel and guide-field conditions.

Our findings support the theory proposed by \cite{li+21a}, indicating the pressure gradient's role in suppressing reconnection outflow.
For sufficiently high temperatures ($\bit \gtrsim 1$), ions demagnetize as they cross the current sheet, leading to heating and increased pressure along the diagonal outflow component of the tensor $P_{ixx}$.
This effect is evident across various guide-field strengths and temperatures. 
Additionally, when $\bit$ falls below 1, the exhaust speed matches the Alfv\'en speed in the presence of a guide field because the ions remain magnetized;
this is in contrast to  the anti-parallel case in which the outflow is always reduced because there is always heating in the exhaust.

The results of this investigation are likely impactful for our understanding of reconnection as a turbulence dissipation mechanism \citep{matthaeus+03, servidio+09, minping+14, haggerty+17, mallet+17, shay+18}.
As the turbulent cascade proceeds, smaller scale reconnection sites are likely to occur with a larger scale mean field,  i.e., with greater guide field strengths.
Additionally, the ongoing dissipation of the turbulent energy will likely have heated the ions undergoing reconnection.
This implies that many smaller scale, guide field reconnection sites may have $\bit \gtrsim 1$, and as such the results of this work is directly applicable for how reconnection dissipates energy in turbulence if the reconnection process has ion coupling.
This effect is more likely to affect hotter turbulent systems such as the intracluster medium (ICM) which has been heated by galaxy cluster mergers \citep{roettiger+96,cavaliere+76}
or for reconnection downstream of any shock wave, where the plasma thermal pressure likely exceeds the magnetic pressure \citep{haggerty+20}, such as in the heliospheric termination shock or in Earth's bow shock, where reconnection has been observed to occur in-situ \citep{gingell+19}.

In summary, this investigation underscores the crucial role of the temperature of the inflowing plasma as a modulator for the heating and outflow dynamics in the case of guide field reconnection. Broader implications are expected for astrophysical phenomena such as collisionless turbulence where guide fields are expected to be important. 
Future research should continue to explore the role of increasing ion temperatures on reconnection dynamics in various astrophysical settings.

\begin{acknowledgments}
The work of CCH was supported by the NSF/DOE Grant PHY-2205991, NSF-FDSS Grant AGS-1936393, NSF-CAREER Grant AGS-2338131; PAC by DoE Grant DE-SC0020294, NASA Grant 80NSSC24K0172, NSF Grant PHY-2308669; MAS by NASA 80NSSC20K019 and NSF AGS-2024198. Simulations were performed on TACC’s Stampede 2 and Purdue’s ANVIL. With allocations through NSF-ACCESS (formally XSEDE) PHY220089 and AST180008.

ACCESS (formally XSEDE)
allocation (TG-AST180008)

\end{acknowledgments}

\appendix
\section{Current Sheet Width}\label{sec:apdx:CS}
\begin{figure}[ht]
    \centering
    \includegraphics[width=0.8\columnwidth]{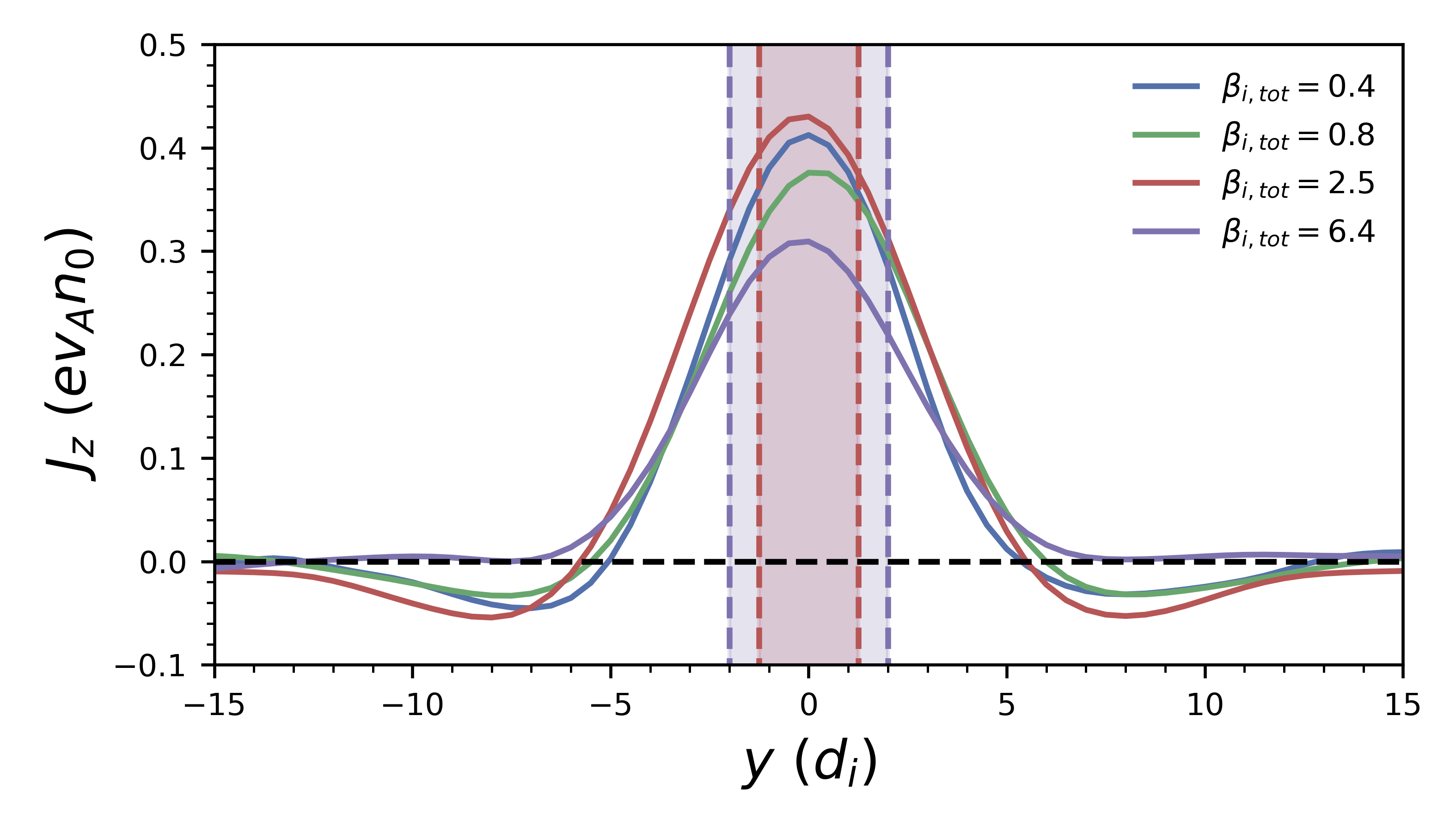}
    \caption{Cuts of $J_z$ through the x-line, along $y$ for reconnection simulations 19, 17, 21, 6 (in order of increasing $\bit$). The simulations correspond to different initial beta values, but a constant guide field strength of $B_g = 2B_r$. The banded purple and red regions show the diameter of the gyro-orbit for the corresponding simulation. As the gyro-radius changes the approximate width of the current sheet stays roughly the same, on the order of a few $d_i$.}
    \label{fig:CS_width}
\end{figure}

In the manuscript, we argue that the outflow component of the ion pressure tensor increases in the large guide field and large $\bit$ regime because of the increasing ion gyro-radius. As the inflowing temperature increases, the gyro-radius becomes larger until its diameter becomes comparable to the length scale over which the magnetic field changes rapidly. We verify this assertion in Fig.~\ref{fig:CS_width}, which shows the out-of-plane current density $J_z$ for 4 different, large guide field simulations ($B_g = 2B_r$). The simulations span from $\bit = 0.4$ to $6.4$, corresponding to the four rightmost blue points in Fig.~\ref{fig:reconn_beta_outflow}. The peak in $J_z$ corresponds to the region where the magnetic field changes most rapidly; the width of the region over which the magnetic field rotates does not change appreciably between simulations. The colored bands show the diameter of a gyro-orbit for their color-corresponding simulations. The red and purple bands correspond to simulations where the outflow speed is reduced, demonstrating that the gyro-radius is comparable to the current sheet thickness.

\bibliographystyle{aasjournal}

\begin{thebibliography}{}
\expandafter\ifx\csname natexlab\endcsname\relax\def\natexlab#1{#1}\fi
\providecommand{\url}[1]{\href{#1}{#1}}
\providecommand{\dodoi}[1]{doi:~\href{http://doi.org/#1}{\nolinkurl{#1}}}
\providecommand{\doeprint}[1]{\href{http://ascl.net/#1}{\nolinkurl{http://ascl.net/#1}}}
\providecommand{\doarXiv}[1]{\href{https://arxiv.org/abs/#1}{\nolinkurl{https://arxiv.org/abs/#1}}}

\bibitem[{{Angelopoulos} {et~al.}(2008){Angelopoulos}, {McFadden}, {Larson},
  {Carlson}, {Mende}, {Frey}, {Phan}, {Sibeck}, {Glassmeier}, {Auster},
  {Donovan}, {Mann}, {Rae}, {Russell}, {Runov}, {Zhou}, \&
  {Kepko}}]{angelopoulos+08}
{Angelopoulos}, V., {McFadden}, J.~P., {Larson}, D., {et~al.} 2008, Science,
  321, 931, \dodoi{10.1126/science.1160495}

\bibitem[{{Caprioli} {et~al.}(2020){Caprioli}, {Haggerty}, \&
  {Blasi}}]{caprioli+20}
{Caprioli}, D., {Haggerty}, C.~C., \& {Blasi}, P. 2020, Submitted to \apj

\bibitem[{{Cavaliere} \& {Fusco-Femiano}(1976)}]{cavaliere+76}
{Cavaliere}, A., \& {Fusco-Femiano}, R. 1976, \aap, 49, 137

\bibitem[{{Dahlin} {et~al.}(2014){Dahlin}, {Drake}, \& {Swisdak}}]{dahlin+14}
{Dahlin}, J.~T., {Drake}, J.~F., \& {Swisdak}, M. 2014, Physics of Plasmas, 21,
  092304, \dodoi{10.1063/1.4894484}

\bibitem[{{Dmitruk} \& {Matthaeus}(2003)}]{matthaeus+03}
{Dmitruk}, P., \& {Matthaeus}, W.~H. 2003, \apj, 597, 1097

\bibitem[{{Drake} {et~al.}(2009{\natexlab{a}}){Drake}, {Cassak}, {Shay},
  {Swisdak}, \& {Quataert}}]{drake+09b}
{Drake}, J.~F., {Cassak}, P.~A., {Shay}, M.~A., {Swisdak}, M., \& {Quataert},
  E. 2009{\natexlab{a}}, \apjl, 700, L16, \dodoi{10.1088/0004-637X/700/1/L16}

\bibitem[{Drake \& Swisdak(2014)}]{drake+14}
Drake, J.~F., \& Swisdak, M. 2014, Physics of Plasmas, 21, 072903,
  \dodoi{10.1063/1.4889871}

\bibitem[{{Drake} {et~al.}(2006){Drake}, {Swisdak}, {Che}, \&
  {Shay}}]{Drake+06}
{Drake}, J.~F., {Swisdak}, M., {Che}, H., \& {Shay}, M.~A. 2006, \nat, 443,
  553, \dodoi{10.1038/nature05116}

\bibitem[{{Drake} {et~al.}(2009{\natexlab{b}}){Drake}, {Swisdak}, {Phan},
  {Cassak}, {Shay}, {Lepri}, {Lin}, {Quataert}, \& {Zurbuchen}}]{drake+09a}
{Drake}, J.~F., {Swisdak}, M., {Phan}, T.~D., {et~al.} 2009{\natexlab{b}},
  Journal of Geophysical Research (Space Physics), 114, A05111,
  \dodoi{10.1029/2008JA013701}

\bibitem[{{Dungey}(1961)}]{dungey61}
{Dungey}, J.~W. 1961, \prl, 6, 47, \dodoi{10.1103/PhysRevLett.6.47}

\bibitem[{{Fermi}(1949)}]{Fermi49}
{Fermi}, E. 1949, Physical Review, 75, 1169, \dodoi{10.1103/PhysRev.75.1169}

\bibitem[{{Gargat{\'e}} {et~al.}(2007){Gargat{\'e}}, {Bingham}, {Fonseca}, \&
  {Silva}}]{gargate+07}
{Gargat{\'e}}, L., {Bingham}, R., {Fonseca}, R.~A., \& {Silva}, L.~O. 2007,
  Comp. Phys. Commun., 176, 419, \dodoi{10.1016/j.cpc.2006.11.013}

\bibitem[{{Gingell} {et~al.}(2019){Gingell}, {Schwartz}, {Eastwood}, {Burch},
  {Ergun}, {Fuselier}, {Gershman}, {Giles}, {Khotyaintsev}, {Lavraud},
  {Lindqvist}, {Paterson}, {Phan}, {Russell}, {Stawarz}, {Strangeway},
  {Torbert}, \& {Wilder}}]{gingell+19}
{Gingell}, I., {Schwartz}, S.~J., {Eastwood}, J.~P., {et~al.} 2019, \grl, 46,
  1177, \dodoi{10.1029/2018GL081804}

\bibitem[{{Haggerty}(2016)}]{Haggerty+16}
{Haggerty}, C.~C. 2016, PhD thesis, University of Delaware

\bibitem[{{Haggerty} \& {Caprioli}(2019)}]{haggerty+19a}
{Haggerty}, C.~C., \& {Caprioli}, D. 2019, \apj, 887, 165,
  \dodoi{10.3847/1538-4357/ab58c8}

\bibitem[{{Haggerty} \& {Caprioli}(2020)}]{haggerty+20}
---. 2020, \apj

\bibitem[{Haggerty {et~al.}(2017)Haggerty, Parashar, Matthaeus, Shay, Yang,
  Wan, Wu, \& Servidio}]{haggerty+17}
Haggerty, C.~C., Parashar, T.~N., Matthaeus, W.~H., {et~al.} 2017, Physics of
  Plasmas, 24, 102308, \dodoi{10.1063/1.5001722}

\bibitem[{Haggerty {et~al.}(2018)Haggerty, Shay, Chasapis, Phan, Drake,
  Malakit, Cassak, \& Kieokaew}]{haggerty+18}
Haggerty, C.~C., Shay, M.~A., Chasapis, A., {et~al.} 2018, Physics of Plasmas,
  25, 102120, \dodoi{10.1063/1.5050530}

\bibitem[{{Haggerty} {et~al.}(2015){Haggerty}, {Shay}, {Drake}, {Phan}, \&
  {McHugh}}]{haggerty+15}
{Haggerty}, C.~C., {Shay}, M.~A., {Drake}, J.~F., {Phan}, T.~D., \& {McHugh},
  C.~T. 2015, \grl, 42, 9657, \dodoi{10.1002/2015GL065961}

\bibitem[{{Hoshino} {et~al.}(2001){Hoshino}, {Hiraide}, \&
  {Mukai}}]{hoshino+01}
{Hoshino}, M., {Hiraide}, K., \& {Mukai}, T. 2001, Earth, Planets and Space,
  53, 627, \dodoi{10.1186/BF03353282}

\bibitem[{{Le} {et~al.}(2012){Le}, {Karimabadi}, {Egedal}, {Roytershteyn}, \&
  {Daughton}}]{le+12}
{Le}, A., {Karimabadi}, H., {Egedal}, J., {Roytershteyn}, V., \& {Daughton}, W.
  2012, Physics of Plasmas, 19, 072120, \dodoi{10.1063/1.4739244}

\bibitem[{Li \& Liu(2021)}]{li+21a}
Li, X., \& Liu, Y.-H. 2021, The Astrophysical Journal, 912, 152,
  \dodoi{10.3847/1538-4357/abf48c}

\bibitem[{Liu {et~al.}(2012)Liu, Drake, \& Swisdak}]{liu+12}
Liu, Y.-H., Drake, J.~F., \& Swisdak, M. 2012, Physics of Plasmas, 19, 022110,
  \dodoi{10.1063/1.3685755}

\bibitem[{Mallet {et~al.}(2017)Mallet, Schekochihin, \& Chandran}]{mallet+17}
Mallet, A., Schekochihin, A.~A., \& Chandran, B.~D. 2017, Journal of Plasma
  Physics, 83, 905830609

\bibitem[{{Parker}(1963)}]{parker63}
{Parker}, E.~N. 1963, \apjs, 8, 177, \dodoi{10.1086/190087}

\bibitem[{{Paschmann} {et~al.}(1979){Paschmann}, {Papamastorakis}, {Sckopke},
  {Haerendel}, {Sonnerup}, {Bame}, {Asbridge}, {Gosling}, {Russel}, \&
  {Elphic}}]{paschmann+79}
{Paschmann}, G., {Papamastorakis}, I., {Sckopke}, N., {et~al.} 1979, \nat, 282,
  243, \dodoi{10.1038/282243a0}

\bibitem[{Phan {et~al.}(2014)Phan, Drake, Shay, Gosling, Paschmann, Eastwood,
  {\O}ieroset, Fujimoto, \& Angelopoulos}]{phan+14}
Phan, T., Drake, J., Shay, M., {et~al.} 2014, Geophysical Research Letters, 41,
  7002

\bibitem[{{Phan} {et~al.}(2013){Phan}, {Shay}, {Gosling}, {Fujimoto}, {Drake},
  {Paschmann}, {Oieroset}, {Eastwood}, \& {Angelopoulos}}]{phan+13}
{Phan}, T.~D., {Shay}, M.~A., {Gosling}, J.~T., {et~al.} 2013, \grl, 40, 4475,
  \dodoi{10.1002/grl.50917}

\bibitem[{{Roettiger} {et~al.}(1996){Roettiger}, {Burns}, \&
  {Loken}}]{roettiger+96}
{Roettiger}, K., {Burns}, J.~O., \& {Loken}, C. 1996, \apj, 473, 651,
  \dodoi{10.1086/178179}

\bibitem[{{Rogers} {et~al.}(2003){Rogers}, {Denton}, \& {Drake}}]{rogers+03}
{Rogers}, B.~N., {Denton}, R.~E., \& {Drake}, J.~F. 2003, Journal of
  Geophysical Research (Space Physics), 108, 1111, \dodoi{10.1029/2002JA009699}

\bibitem[{{Rogers} {et~al.}(2001){Rogers}, {Denton}, {Drake}, \&
  {Shay}}]{rogers+01}
{Rogers}, B.~N., {Denton}, R.~E., {Drake}, J.~F., \& {Shay}, M.~A. 2001, \prl,
  87, 195004, \dodoi{10.1103/PhysRevLett.87.195004}

\bibitem[{{Rowan} {et~al.}(2017){Rowan}, {Sironi}, \& {Narayan}}]{rowan+17}
{Rowan}, M.~E., {Sironi}, L., \& {Narayan}, R. 2017, \apj, 850, 29,
  \dodoi{10.3847/1538-4357/aa9380}

\bibitem[{Sato \& Hayashi(1979)}]{sato+79}
Sato, T., \& Hayashi, T. 1979, Phys. Fluids, 22, 1189–,
  \dodoi{https://doi.org/10.1063/1.862721}

\bibitem[{{Servidio} {et~al.}(2009){Servidio}, {Matthaeus}, {Shay}, {Cassak},
  \& {Dmitruk}}]{servidio+09}
{Servidio}, S., {Matthaeus}, W.~H., {Shay}, M.~A., {Cassak}, P.~A., \&
  {Dmitruk}, P. 2009, Phys. Rev. Lett., 102, 115003,
  \dodoi{10.1103/PhysRevLett.102.115003}

\bibitem[{{Sharma Pyakurel} {et~al.}(2019){Sharma Pyakurel}, {Shay}, {Phan},
  {Matthaeus}, {Drake}, {TenBarge}, {Haggerty}, {Klein}, {Cassak}, {Parashar},
  {Swisdak}, \& {Chasapis}}]{sharma+19}
{Sharma Pyakurel}, P., {Shay}, M.~A., {Phan}, T.~D., {et~al.} 2019, Physics of
  Plasmas, 26, 082307, \dodoi{10.1063/1.5090403}

\bibitem[{Shay {et~al.}(1999)Shay, Drake, Rogers, \& Denton}]{shay+99}
Shay, M., Drake, J., Rogers, B., \& Denton, R. 1999, Geophysical Research
  Letters, 26, 2163

\bibitem[{{Shay} {et~al.}(2018){Shay}, {Haggerty}, {Matthaeus}, {Parashar},
  {Wan}, \& {Wu}}]{shay+18}
{Shay}, M.~A., {Haggerty}, C.~C., {Matthaeus}, W.~H., {et~al.} 2018, Physics of
  Plasmas, 25, 012304, \dodoi{10.1063/1.4993423}

\bibitem[{{Shay} {et~al.}(2014){Shay}, {Haggerty}, {Phan}, {Drake}, {Cassak},
  {Wu}, {Oieroset}, {Swisdak}, \& {Malakit}}]{shay+14}
{Shay}, M.~A., {Haggerty}, C.~C., {Phan}, T.~D., {et~al.} 2014, Physics of
  Plasmas, 21, 122902, \dodoi{10.1063/1.4904203}

\bibitem[{Sonnerup(1981)}]{sonnerup+81}
Sonnerup, B. U.~. 1981, Journal of Geophysical Research (Space Physics), 86,
  10049, \dodoi{https://doi.org/10.1029/JA086iA12p10049}

\bibitem[{{TenBarge} {et~al.}(2014){TenBarge}, {Daughton}, {Karimabadi},
  {Howes}, \& {Dorland}}]{tenbarge+14}
{TenBarge}, J.~M., {Daughton}, W., {Karimabadi}, H., {Howes}, G.~G., \&
  {Dorland}, W. 2014, Physics of Plasmas, 21, 020708, \dodoi{10.1063/1.4867068}

\bibitem[{{Wan} {et~al.}(2014){Wan}, {Rappazzo}, {Matthaeus}, {Servidio}, \&
  S.}]{minping+14}
{Wan}, M., {Rappazzo}, A.~F., {Matthaeus}, W.~H., {Servidio}, S., \& S., O.
  2014, \apj, 797, \dodoi{10.1088/0004-637X/797/1/63}

\bibitem[{{Yamada} {et~al.}(1997){Yamada}, {Ji}, {Hsu}, {Carter}, {Kulsrud},
  {Bretz}, {Jobes}, {Ono}, \& {Perkins}}]{yamada+97}
{Yamada}, M., {Ji}, H., {Hsu}, S., {et~al.} 1997, Physics of Plasmas, 4, 1936,
  \dodoi{10.1063/1.872336}

\bibitem[{{Yoo} {et~al.}(2014){Yoo}, {Yamada}, {Ji}, {Jara-Almonte}, \&
  {Myers}}]{yoo+14}
{Yoo}, J., {Yamada}, M., {Ji}, H., {Jara-Almonte}, J., \& {Myers}, C.~E. 2014,
  Physics of Plasmas, 21, 055706, \dodoi{10.1063/1.4874331}

\bibitem[{Zhang {et~al.}(2021)Zhang, Guo, Daughton, Li, \& Li}]{zhang+21}
Zhang, Q., Guo, F., Daughton, W., Li, H., \& Li, X. 2021, Physical Review
  Letters, 127, 185101

\bibitem[{Zweibel \& Yamada(2009)}]{zweibel+09}
Zweibel, E.~G., \& Yamada, M. 2009, Annual Review of Astronomy and
  Astrophysics, 47, 291, \dodoi{10.1146/annurev-astro-082708-101726}

\end{thebibliography}

\end{document}